# Evidence that Threatening Situations Enhance Creativity

**Sean N. Riley (research@seanriley.ca)**
Department of Psychology, University of British Columbia
3333 University Way, Kelowna B.C V1V 1V7

**Liane Gabora (liane.gabora@ubc.ca)**
Department of Psychology, University of British Columbia
3333 University Way, Kelowna B.C V1V 1V7

## Abstract

We tested the hypothesis that threatening situations enhance creativity. 60 participants viewed a series of photographs and rated them on level of threat. They then wrote two short stories: one based on the photograph they rated as most threatening, and the other based on the photograph they rated as least threatening. The stories were rated for level of creativity. Paired samples t-testså revealed that stories based on threatening pictures produced a higher degree of creativity than those based on non-threatening pictures. Theoretical frameworks consistent with these findings are discussed.

## Introduction

Creativity can be defined as the ability to generate ideas, interpretations, or solutions that are both novel, and meaningful or appropriate (Sternberg *et al.,* 2010). Creativity is widely associated with personal fulfillment (May, 1975; Rogers, 1959), self-actualization (Maslow, 1959), and with maintaining a competitive edge in the marketplace. Creative therapies are useful in clinical contexts, including the assessment and resolution of conflict (Goldblatt *et al.,* 2011), dementia (Hannermann, 2006), self-esteem (Anzules, Haennl & Golay, 2007), and stress reduction (Curl, 2008). In addition, creative individuals tend to have higher levels of life satisfaction (Tan, Ho, Ho, & Ow, 2008), emotional intelligence (Noferesti & Al-ghorabaie, 2011), and intelligence in general (Batey & Furnham, 2009). These findings suggest that creativity is, generally speaking, a positive attribute with numerous constructive byproducts.

However, there are significant drawbacks to creativity (Cropley, Cropley, Kaufman, & Runco, 2010; Ludwig, 1995). Generating creative ideas is time consuming, and a creative solution to one problem often generates other problems, or has unexpected negative side effects that may only become apparent after much effort has been invested. Creative people often reinvent the wheel, and may be more likely to bend rules, break laws, and provoke social unrest (Cropley, Kaufman, & Cropley, 2003; Sternberg & Lubart, 1995; Sulloway, 1996). They tend to be more emotionally unstable and prone to affective disorders such as depression and bipolar disorder, and have a higher incidence of schizophrenic tendencies than other segments of the population (Andreason, 1987; Flaherty, 2005; F. Goodwin & Jamison, 1990). Computational models suggest that there is a detrimental impact on society if either the ratio of creative to relatively uncreative individuals is too high, or if the creative individuals are *too* creative (Gabora & Firouzi, 2012; Gabora & Leijnen, 2009; Leijnen & Gabora, 2009).

There is also preliminary evidence that situations that are demanding, threatening, or involve conflict, put one in a more creative state of mind. For example, it has been shown that individuals who are in the midst of conflict set broader and more inclusive cognitive categories (De Dru, Carsten & Nijstad, 2008). Creativity is positively correlated with aggression (Tacher & Readdick, 2006), group conflict (Troyer & Youngreen, 2009) anxiety (Carlsson, 2002), and dishonesty (Gino & Ariely, 2011). Finally it has been found that negative affect leads to greater creative output (Akinola, Mendes, 2008).

This study further investigates the hypothesis that threatening situations put one in a more creative state of mind. Specifically, it assesses whether stories written in response to threatening stimuli are more creative than stories written in response to non-threatening stimuli.

## The Study

### Participants

Participants ($n$ = 60; 19 M, 41 F) were recruited for this study through the University of British Columbia (Okanagan campus) course credit for research participation program used in introductory psychology classes. Participants received 1% added to their overall course grade.

### Stimuli

Participants were shown 15 photographs depicting situations that had been independently classified as either threatening or nonthreatening by both experimenters. Classification was determined by assessing whether a normal individual would feel significantly at risk of harm or death if they were in the scenario depicted by the photograph. The photographs classified as 'threatening' were deemed to be threatening not directly, but indirectly, in the same sense that one feels threatened by a war movie or horror film.

Examples of the threatening and non-threatening photographs used to generate stories in this study are given in Figures 1 and 2.

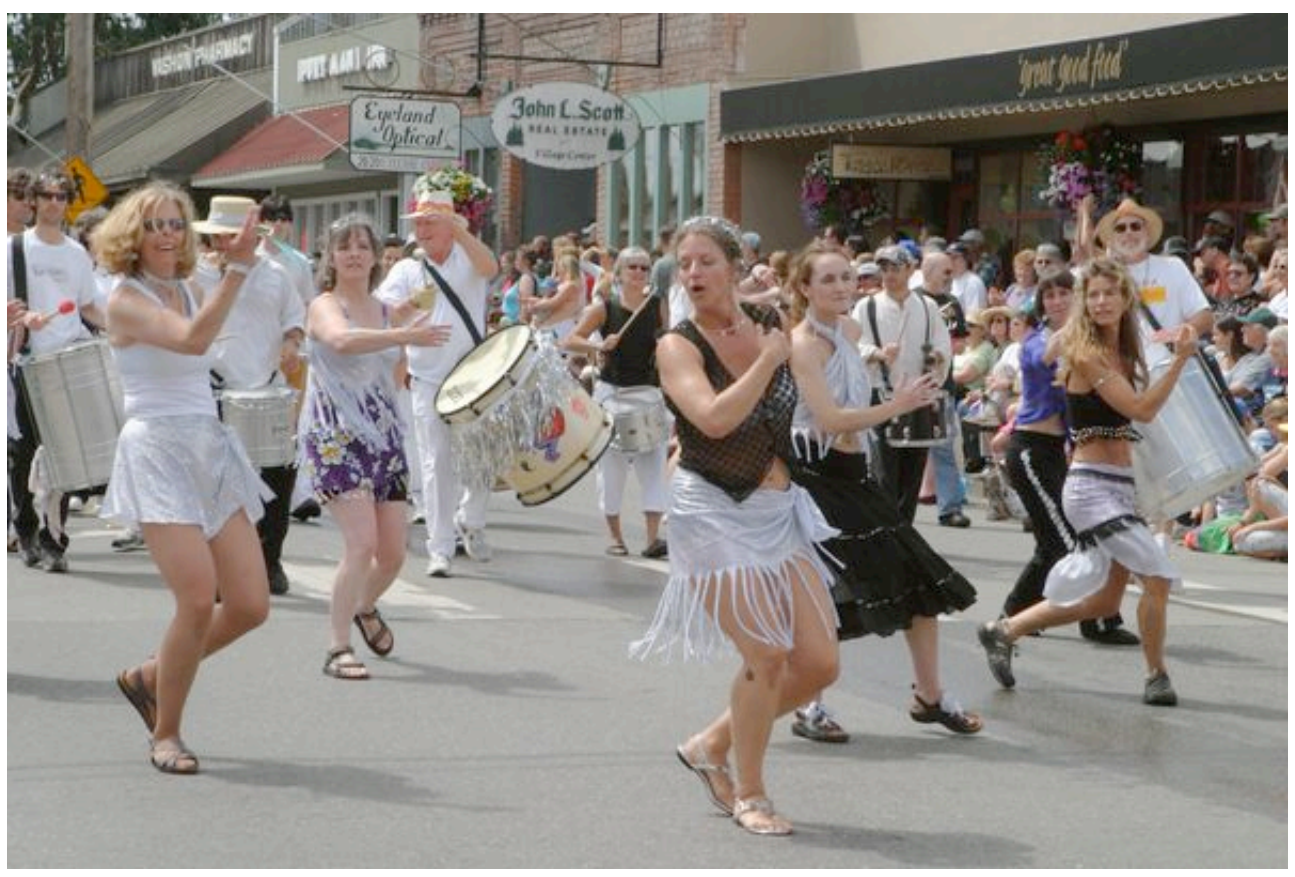

Figure 1. Example of non-threatening photograph.

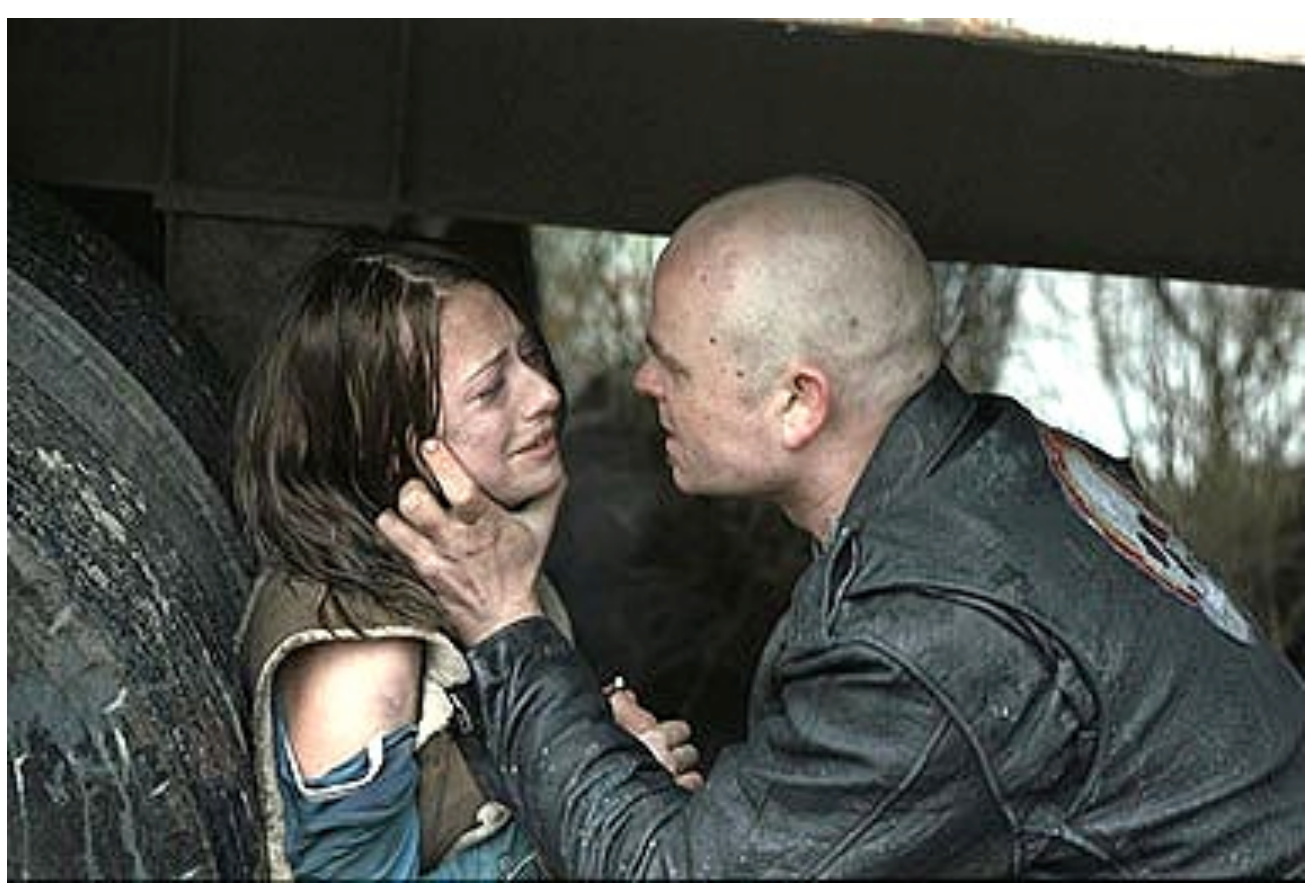

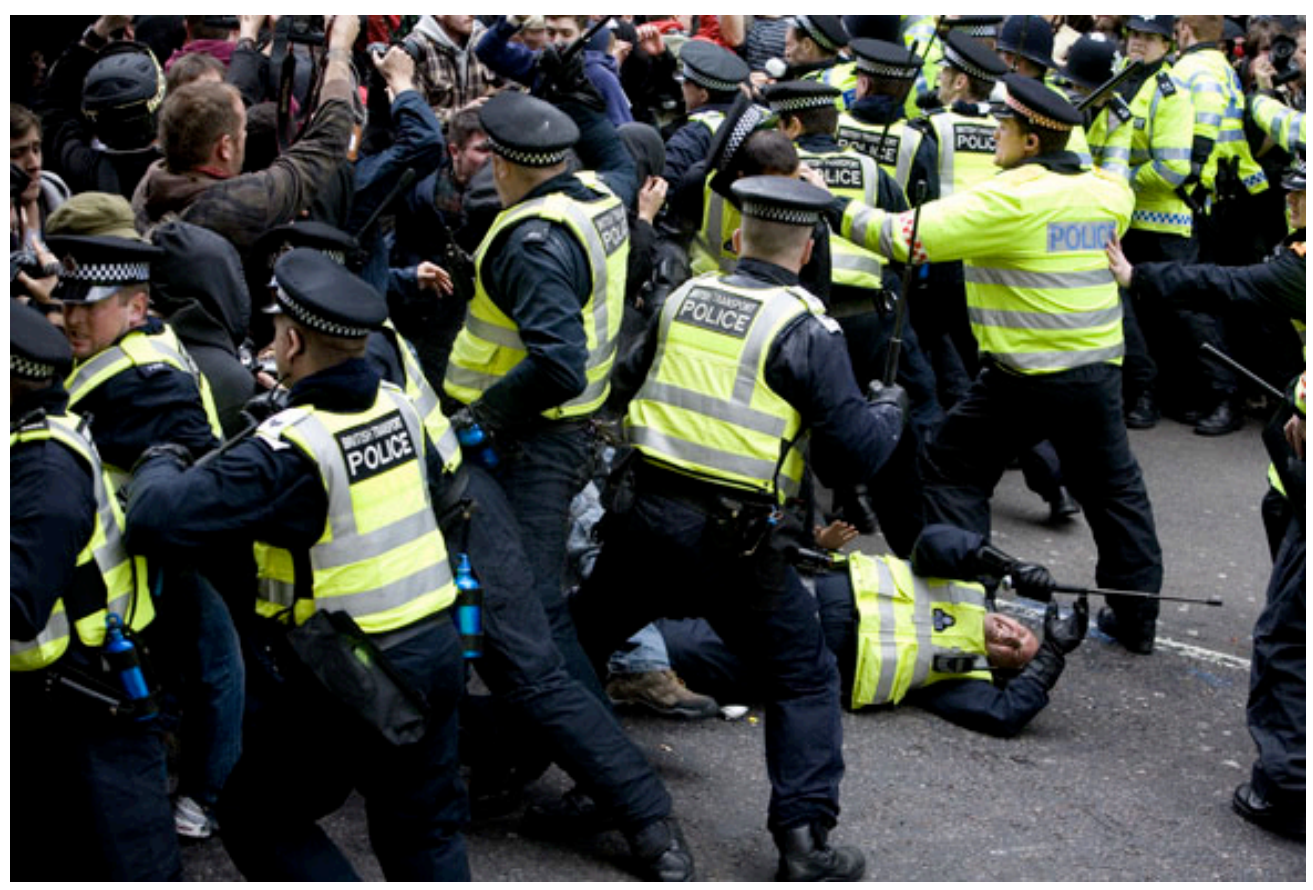

Figure 2. Two examples of threatening photographs.

## Procedure

The participants were asked to rate each photograph according to how threatening they believed it to be on a 7-point Likert scale. Each participant was then given 30 minutes to write two short stories: one about the photograph that was rated by that participant as most threatening, and one about the photograph that was rated by that participant as least threatening. If two photographs were rated as equally threatening, the photograph that was presented first in the sequence was used. To guard against order effects, both story order and picture order were counterbalanced.

## Raters

Four raters were University of British Columbia undergraduates who were enrolled in an advanced psychology research methods and statistics course. They are referred to as *student raters*. They consisted of three females and one male.

The fifth rater was a well-known and extensively published fiction author who was not reimbursed for his participation. He is referred to as the *expert rater*. He was male.

## Rating

Raters were trained to evaluate participant responses using a previously developed rubric for assessing story creativity, the 'Creativity' portion of the Wisdom Intelligence Creativity Synthesized (WICS) rubric (Sternberg, 2005; Sternberg *et al*., 2010, 2012), and a set of stories that had previously established creativity ratings. The WICS rubric assesses story creativity on the basis of evidence of the respondent`s ability to provide novel yet meaningful ideas, narratives or interpretations of situations, or solutions to problems, or to view situations and events in a new and meaningful way.

Raters practiced on the sample stories until their ratings on these sample stories correlated significantly with these stories' previously established creativity ratings.

# Results

The creativity ratings provided by the expert rater were significantly positively correlated with the creativity ratings provided by the student raters, as shown in Table 1.

Table 1: Bivariate correlation between expert rater and each of the four student raters.

| | **Rater One** | **Rater Two** | **Rater Three** | **Rater Four** |
|---|---|---|---|---|
| Non-threatening | .289* | .454** | .286* | .313* |
| Threatening | .376** | .156 | .368** | .477** |

*Significant at < .05; **Significant at < .01.

Intraclass correlation was used to assess the level of agreement amongst all five raters. They were significantly in agreement, as shown in Table 2.

Table 2. Inter-class correlation between all raters.

We created two rater groups: the average of all the raters' scores, as well as the average of the two most central raters, for each of the stories. The central raters were selected by dropping the raters with the highest and lowest means, and then selecting from the remaining three raters the two whose mean ratings were closest to one another. We conducted a paired samples t-test between stories based on threatening pictures, and stories based on non-threatening pictures, for both rater groups. The mean creativity scores for stories generated in response to non-threatening and threatening stimuli are shown in Figure 4.

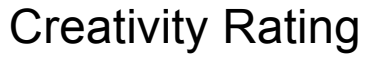

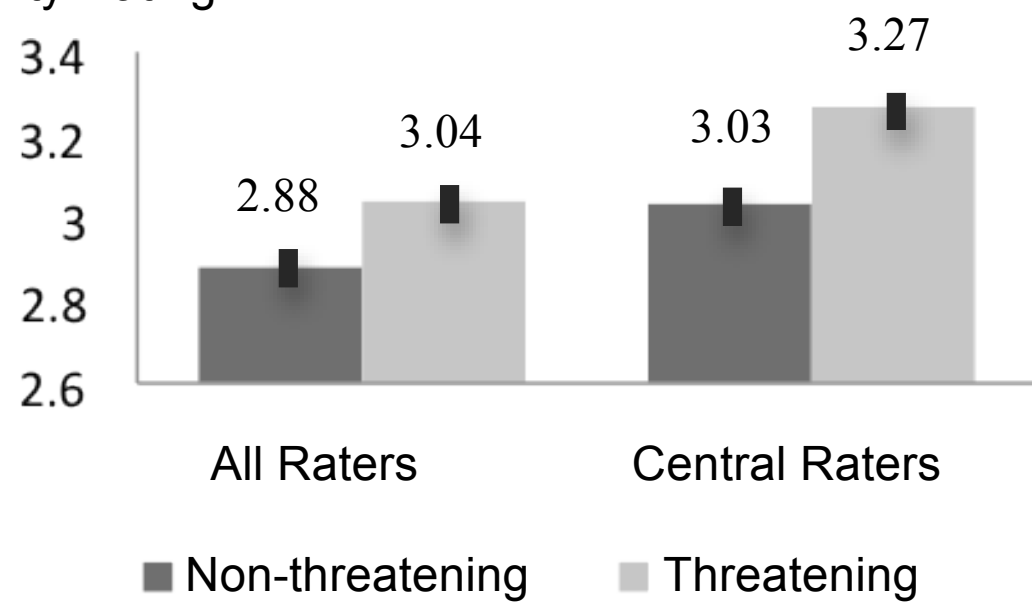

Figure 4. Mean creativity scores for stories generated in response to non-threatening and threatening stimuli.

The results indicate a high level of agreement between raters ($p$ = .001) and support for our hypothesis. As shown in Table 3, using both the average of all raters, as well as the two central raters, paired sample t-tests revealed that stories based on threatening pictures were rated as significantly more creative than those based on non-threatening pictures ($p < .05$).

Table 3. Paired samples t-tests.

| | t-value | SE | Cohen's $d$ | Sig. |
|---|---|---|---|---|
| All Raters | 2.08 | .077 | .269 | .041 |
| Central Raters | 2.29 | .101 | .296 | .026 |

## Discussion

These results support our hypothesis that stories written in response to threatening photographs elicit a higher degree of creativity than those written in response to non-threatening photographs. Several theoretical frameworks are consistent with and could help explain these findings.

One kind of explanatory framework comes from creativity research. It is widely assumed that the creative process involves *searching* through memory and/or *selecting* amongst a set of predefined candidate ideas. For example, computer scientists have modeled the creative process as a heuristic search (e.g., Simon, 1973, 1986). In psychology, there is much evidence for, and discussion of, the role of *divergent thinking* in creativity (Guilford, 1968; for a review see Runco, 2010). Divergent thinking is presumed to involve the generation of multiple, often unconventional, possibilities. When construed from this perspective, the creative process often goes hand-in-hand with the notion of selection, since if you come up with multiple alternatives you eventually weed some of them out. Indeed, many well-known theories of creativity, such as the Geneplore model (Finke, Ward, & Smith, 1992), and the Darwinian theory of creativity (Simonton, 1999) involve two stages: the generation of possibilities, followed by the exploration and ultimately selective retention of the most promising of them.

| | ICC | Value | Sig. |
|---|---|---|---|
| Single Measures | .430 | 8.542 | .001 |
| Average Measures | .883 | 8.542 | .001 |

However, there are strong neurobiological (Gabora, 2010a), experimental (Gabora, 2010b; Gabora, O'Connor & Ranjan, in press), and theoretical (Gabora, 2005, 2007), reasons to believe that the generation stage of creative thinking may be *divergent* not in that it moves in multiple directions or generates multiple possibilities, but in the sense that it produces a raw idea that is vague or unfocused, and that requires further processing to become viable. Similarly, the exploration stage of creative thinking may be *convergent*, not in ordinary sense that it entails selecting from amongst alternatives, but in the sense that it entails considering a vague idea from different perspectives until it comes into newly defined focus. In other words, the terms 'divergent' and 'convergent' may be applicable to creative thought in the sense of going from well-defined to ill-defined, and *vice versa*. Although a particular creative thinking process *may* involve search or selection amongst multiple well-defined possibilities, it *need* not, and moreover, that *selection* need not figure prominently in a general theory of creativity.

According to this alternative view of creativity, creative individuals wrestle with those issues or ideas that are, for them, in a state of *potentiality*. An artist might wrestle with how to capture the feeling of a particular landscape, and a writer might wrestle with depicting how events in an imaginary world would unfold. Over time, creative ideas come to assume a form that is more fully *actualized,* or well-defined, as they are considered from different perspectives in accordance with the constraints of the domain in which they are expressed. By giving form to that which exists in a state of potentiality, the individual gains a richer understanding and appreciation of it, as well as a sense of control or mastery over it.

Central to this view of creativity is the notion of a *worldview*. The term worldview is used to refer to one's internal model of the world, as well as one's values, predispositions, and habitual patterns of response (Gabora, 2000, 2008; Gabora & Aerts, 2009). Each idea the creator comes up with is construed as a different expression of the same underlying core network of understandings, beliefs, and attitudes. Thus an individual's outputs are inter-related, and potentially pave the way for one another. By adopting the notion of a worldview, our account places equal emphasis on external creative outcomes and the internal cognitive and emotional restructuring brought about by the creative process. The cognitive reorganization and personality dynamics (*e.g.,* involving wellbeing, self-discipline, or self-discovery) are viewed as the internal, less readily measurable, but equally important, counterpart to external manifestations of the creative process. The transformation that occurs on canvas or on the written page is mirrored by a sense of personal, cognitive transformation and self-discovery from within.

In the case of our study, this alternative theory suggests that the threatening stimuli created a dichotomy in terms of one's understanding of the world. On the one hand, we believe that the world is a just and fair world, and that we, as individuals, are deserving of just and fair treatment. However, the threatening stimulus confronts this conception by imposing upon us a negative reality that threatens our internal model of how the world operates. In response to this, we tap into our creative potential and hone in on a suitable explanation for the threat's existence. This is done in an attempt to reconcile the worldview dichotomy, and impose a sense of meaning and understanding as to why this negative reality exists, ultimately forging a new and cohesive worldview structure.

These findings are consistent with the previously mentioned literature on the dark side of creativity. The positive correlation between negative affect and creativity (Akinola & Mendes, 2008) gives credence to the notion that creativity can arise from worldview disequilibrium, that is, from a schism between one's reality and one's internal conception of reality. In the case of negative affect, one's reality is that one is experiencing unpleasant and unsettling emotions; however, the internal conception of reality is predicated on the notion that one should be happy. In response to this dichotomy, the creative process helps to create a new, cohesive worldview structure that reduces negative affect (*e.g.,* Curl, 2008).

These findings can also be interpreted in terms of the cognitive-tuning hypothesis (Schwarz, 2002). This theory posits that our mood mirrors our surroundings, and mood-states in turn, affect how information from the environment gets processed. In the case of a negative mood-state, one narrows attention and focuses on the negative environment. One then becomes more motivated to appropriately cope with the negative mood, and engage in bottom-up processing of the environment to understand the factors underpinning the negative environment. Conversely, a positive-mood state does not require an in-depth assessment of the underpinning factors; thus, one engages in top-down processing, and uses heuristics to understand the environment. In terms of creative output, the bottom-up processing used in negative mood-states allows for a more in-depth analysis of the environment, which can lead to more novel interpretations of its underpinning factors. The positive-mood state does not necessitate this in-depth analysis; thus a common, less novel narrative is developed. In the study reported here, the threatening photographs may have induced a negative mood-state, and the non-threatening photographs may have induced a neutral or positive mood-state, with the ensuing impact on information processing and creative output.

To best of our knowledge, this paper is the first study to directly assess the role of threatening stimuli on creative output. However, the study has some weaknesses and limitations. First, the photographs used were not drawn from the International Affective Picture System. Although the photographs used in the study did, on a qualitative level, hold a high degree of consistency in terms of arousal and valence, future studies should consider using standardized photographs. Furthermore, due to time constraints, participants were limited to 15 minutes of writing time per story. It is possible that, given adequate time, we could see a convergence of creative output. Despite these limitations, we believe the study yielded some highly intriguing preliminary findings concerning the role of threat on creative output, which pave the way for further research into this phenomenon.

## Acknowledgments

We would like to acknowledge grants to the second author from the National Science and Engineering Research Council of Canada, and the Fund for Scientific Research of Flanders, Belgium.

# Appendices

Appendices A, B, C, and D provide examples of stories that scored high or low with respect to creativity, generated in response to non-threatening or threatening photographs. (The stories may end abruptly because they were asked to stop at the end of the allotted writing time.)

## Appendix A

The following is an example of a high scoring story produced in response to the non-threatening stimulus shown in Figure 1.

In Jaredsville, Fourth of July parade was something everyone in town looked forward to. The sleepy town in Missouri came together once a year to provide food, movies, festivities, and friendship to their fellow townsman. The streets are filled with people, some searing food on the grill while others paint children's faces into tigers and clowns.

But everyone came, of course, to see the Jaredsville professional dance exhibition. A few of the ladies in town had once been professional dancers, traveling across the world displaying their beauty and raw skills. They all began to slowly get in position with the start of the drums. The seguins on their clothes danced in the sunlight to the movement of their bodies. The drums began beating faster and faster and the women expertly kept with the beat, dancing faster and more radically until their bodies looked like surges of light, The effect was almost hypnotizing as the crowd had quieted down; became fixated by the dancing flashes of light.

And just as quickly as they had started, they finished. As they left the middle of the street, the crowd applauded them fiercely. Yet, something had changed; the crowd was no longer the large bustling rabble it once was. It had now been neutralized, as if soother by the dancers.

## Appendix B

The following is an example of a low scoring story produced in response to the non-threatening stimulus shown in Figure 1.

The town had a celebration today, an annual celebration that celebrated the culture and music of the country. The women were the center of the attention, they dressed in black or white like the men did, but they were skirts. The women danced to the music and paraded around the whole town. Some knew exactly what was going on and the moves to perform such a dance. Others had their eyes towards the 'leader', who perhaps was the one who made the routine.

The bright sunny day made it all the more amazing to watch the show, as old men were wearing matching hats and marching around with drums. There were people everywhere along the sides of the road, up in the balcony, just to get a glimpse of the festivities that were happening. Some cheered and waved around, and some took pictures. All of the stores in that main road seemed to be closed, probably because of how hectic and wild the place was getting!

It probably was a great success; the warm weather, the loud music, and cheering of people made the day so great! This celebration will most likely continue with similar events every year. After all, it is a celebration of culture and music to the country!

## Appendix C

The following is an example of a high scoring story produced in response to the threatening stimulus shown in Figure 3.

The man pushed Cheryl under the truck, "Where is he?" he was screaming at her. Cheryl could only sob more, trying to control the emotion was impossible, an endless waterfall of emotions falling over her was all she could feel. "I said, where is he!" The unknown man yelling once again, for the first time in days she felt something, the clasp of his hand over her face. Only days earlier someone she loved had placed his hands in the same place, but this time there was no tenderness, no caress to show they cared. "I will put your body under the tire and back over you slowly, crushing one bone at a time 'till you tell me where the man I am looking for is!" She looked down still focusing on the hand on her face, how she missed him! A few days ago, her lover had left. Where, she had no idea, "Business to take care of" he had said, but what business would leave her alone so this man, the one with the strangling hold on her face, could find her. She had tried to run, but with everyone here who only cared for themselves (all they could care for really with so much sickness and death around), no one answered her screams, her frantic running and trying to escape this man. She riped her shirt on one of the branches she hit and had the sensation of a bruise forming. She paid no heed to it, only a small discomfort when she compared it to the gaping hole she felt in her chest. Suddenly the man's coat began vibrating, he answered the phone with his hand still around her neck. "Mmmm ok, yes, dead? Good." Without warning he dropped her to the ground and stood looking over her. "Turns out we don't need you" he said leering, "but I do know some of the boys may want you". He picked her up by the hair and began dragging her to the truck "And who knows? Maybe we will let you see your man's body before we lay you down for us" Cheryl tried to force herself once more, with less gusto than before as a part of her heart was crying out what the bald man said to be true was not! She would believe he was alive until she laid her head on his still, dead chest and felt him one last time.

## Appendix D

The following is an example of a low scoring story produced in response to the threatening stimulus shown in Figure 3.

His hands were cold and gritty upon my face they began to turn white as he squeezed harder and threatened me for the money. All I wanted to do was run away, run away from this life in general. He looked at me with such distain. I wonder how I even got to this point in my life. My heart raced while tears poured down my face, he was a vile man with the intent to kill. "Please" I repeated over and over, "give me one more chance. I'll get you your money." He spat on the ground under the bridge at my feet like I was a piece of scum. I wanted a different life but this grease bag was what I had to deal with every day. Constantly abusing every inch of me. The only thing that scared me more than staying with him was leaving him. He'd find me, he always did. I didn't just need to break away physically, but emotionally and for my own safety. The numbness of the drug he provided was my only gateway out of this hell hole. This could have been the last time I ever saw daylight. Daylight and his eyes, burning with power over every meek creature that would obey him.